\documentclass[preprint,aps]{revtex4}

\usepackage{graphics}
\usepackage{latexsym,amssymb}
\usepackage{bm}
\begin{document}

\preprint{APS}
\title{ Chaotic Behavior of Positronium\\
       in a Magnetic Field.}

\author{J.L. Anderson}
\email{jlanders@stevens.edu}
\author{R.K. Murawski} 
\email{rmurawsk@stevens.edu}
\author{G.Schmidt}
\email{gschmidt@stevens.edu}

\affiliation{Department of Physics and Engineering Physics\\ 
        Stevens Institute of Technology\\
        Hoboken, N.J. 07030}

\date{\today}

\begin{abstract}

   Classical motion of positronium embedded in a magnetic field is studied, and
   computation shows the emergence of chaotic orbits. Recent work investigating
   quantum behavior of this system predicts extremely long lifetimes \cite{jA97}
   \cite{jS98}. Chaos assisted tunneling however may lead to significant
   shortening of the lifetime of this system.
\end{abstract}

%\keywords{positronium,chaos,magnetic field}

\maketitle

Two interacting charged particles placed in a magnetic field
exhibit chaotic motion. This has been studied for the hydrogen and Rydberg
atom \cite{jD84} and the scattering of electrons on positive nuclei
\cite{gS00}.This has an impact on the electrical conductivity of fully
ionized plasmas \cite{gS02} \and \cite{bH02}.

Here we study the classical motion of positronium in a magnetic field. In
the absence of a magnetic field the positronium has a very short lifetime.
It was found by Ackermann et. al.\cite{jA97} that in a strong
magnetic field the positronium can have an extremely long lifetime ''up to
the order of one year''\cite{jA97}.

We find that the classical motion is chaotic, which usually leads to chaos
assisted tunneling \cite{pt01} which should significantly reduce the
lifetime of this system.

The calculation includes the case of crossed electric and magnetic fields,
provided that the ratio of the field strengths $E/B$ does not exceed the
speed of light. In this case the electric field can be eliminated by a
Lorentz transformation.

The motion of two particles with charges \ +e and \ -e of equal
mass m moving, in a uniform magnetic field $\ \mathbf{B}$ are described by
the equations

\begin{equation}\label{E:n1}
{m}\bm{\ddot{r}}_{1}=\allowbreak e\bm{\dot{r}}_{1}\times 
\bm{B}-\frac{e^{2}}{4\pi \epsilon_{0}}\frac{\bm{r}_{1}-\bm{r}
_{2}}{\mid \bm{r}_{1}-\bm{r}_{2}\mid ^{3}}
\end{equation}

\begin{equation}\label{E:n2}
{m}\bm{\ddot{r}}_{2}=-e\bm{\dot{r}}_{2}\times \bm{B}+\frac{e^{2}}{
4\pi \epsilon_{0}}\frac{\bm{r}_{1}-\bm{r}_{2}}{\mid \bm{r}
_{1}-\bm{r}_{2}\mid ^{3}}
\end{equation}

Adding (\ref{E:n1}) and (\ref{E:n2}) gives

\begin{equation}\label{E:1p2}
{m}(\bm{\ddot{r}}_{1}+\bm{\ddot{r}}_{2})=e(\bm{\dot{r}}_{1}-
\bm{\dot{r}}_{2})\times \bm{B}
\end{equation}

Introducing the new coordinates

\begin{equation}\label{E:br}
\bm{r}_{1}+\bm{r}_{2}=\bm{R} 
\end{equation}

\begin{equation}\label{E:lr}
\bm{r}_{1}-\bm{r}_{2}=\bm{r} 
\end{equation}

and integrating Eq. (\ref{E:1p2}) results in

\begin{equation}\label{E:alpha}
{m}\bm{\dot{R}}-e\bm{r}\times \bm{B}=\boldsymbol{\alpha }
\end{equation}
 
where $\boldsymbol{\alpha }$ is a constant vector. Subtracting Eq.(\ref{E:n2}) from
(\ref{E:n1}) gives

\begin{equation}\label{E:2m1}
{m}\bm{\ddot{r}}=e\bm{\dot{R}}\times \bm{B}-\frac{e^{2}\bm{r}
}{2\pi \epsilon _{0}r^{3}}
\end{equation}

and using Eq.(\ref{E:alpha})

\begin{equation}\label{E:8}
{m}\bm{\ddot{r}}=\frac{e}{m}(e\bm{r}\times \bm{B}+\boldsymbol{\alpha }
)\times \bm{B}-\frac{e^{2}}{2\pi \epsilon _{0}}\frac{\bm{r}}{r^{3}}
\end{equation}

Introducing  the cyclotron frequency $\ \omega _{c}=eB/m,$ and  choosing $
\bm{B}$ pointing in the z direction $\bm{B}=B\bm{e}_{3},$ Eqs.
(\ref{E:8}) and (\ref{E:alpha}) become

\begin{equation}\label{E:9}
\bm{\ddot{r}}/\omega _{c}^{2}=(\bm{r}\times \bm{e}_{3})\times 
\bm{e}_{3}+\boldsymbol{\alpha }\times \bm{e}_{3}/(eB)-\frac{m}{2\pi
\epsilon _{0}B{{}^{2}}}\frac{\bm{r}}{r{{}^{3}}}
\end{equation}

\begin{equation}\label{E:10}
\bm{\dot{R}}/\omega _{c}-\bm{r}\times \bm{e}_{3}=\boldsymbol{\alpha 
}/eB
\end{equation}

With the dimensionless variables $\ \omega _{c}t\rightarrow t$ and \ $r(2\pi
\epsilon_{0}B{{}^2}/m)^{1/3}\rightarrow r,$ one arrives to the
dimensionless equations of motion

\begin{equation}\label{E:11}
\bm{\ddot{r}}=(\bm{r}\times \bm{e}_{3})\times \bm{e}_{3}+%
\boldsymbol{\alpha}^{'}\times \bm{e}_{3}-\bm{r}/r{{}^{3}}
\end{equation}

\begin{equation}\label{E:12}
\bm{\dot{R}}-\bm{r}\times \bm{e}_{3}=\boldsymbol{\alpha}^{'}
\end{equation}

where $\ \boldsymbol{\alpha}^{'}=\boldsymbol{\alpha }/eB$ is the dimensionless constant
vector. Since $\bm{r}\times \bm{e}_{3}$ has no component in the z
direction, it is convenient to chose a coordinate system where the initial
value \ of $\dot{R}_{z}(0)=0$, so the constant $\ \boldsymbol{\alpha}^{'}$ 
is a vector in the $x-y$ plane, $\bm{\alpha^{'}}=a\bm{e}%
_{1}+b\bm{e}_{2_{{}}}.$Without loss of generality one may chose either
a=0 or b=0. So Eq.(\ref{E:11}) \ becomes

\begin{equation}\label{E:13}
\bm{\ddot{r}}+\bm{r}_{\perp }=-\bm{r}/r{{}^{3}}+b\bm{e}_{1}
\end{equation}

where $\bot $ means perpendicular to the z axis.\\

First we study the two dimensional case

\begin{equation}\label{E:14}
 \begin{array}{l}
  \ddot{x}+x=-x/r{{}^{3}}+b\\
  \ddot{y}+y=-y/r{{}^{3}}
  \end{array}
\end{equation}

with the Hamiltonian

\begin{equation}\label{E:15}
H=p{{}^{2}}/2+r{{}^{2}}/2-1/r-bx
\end{equation}

\begin{figure}[tbh]
\vspace{1.in}
\includegraphics{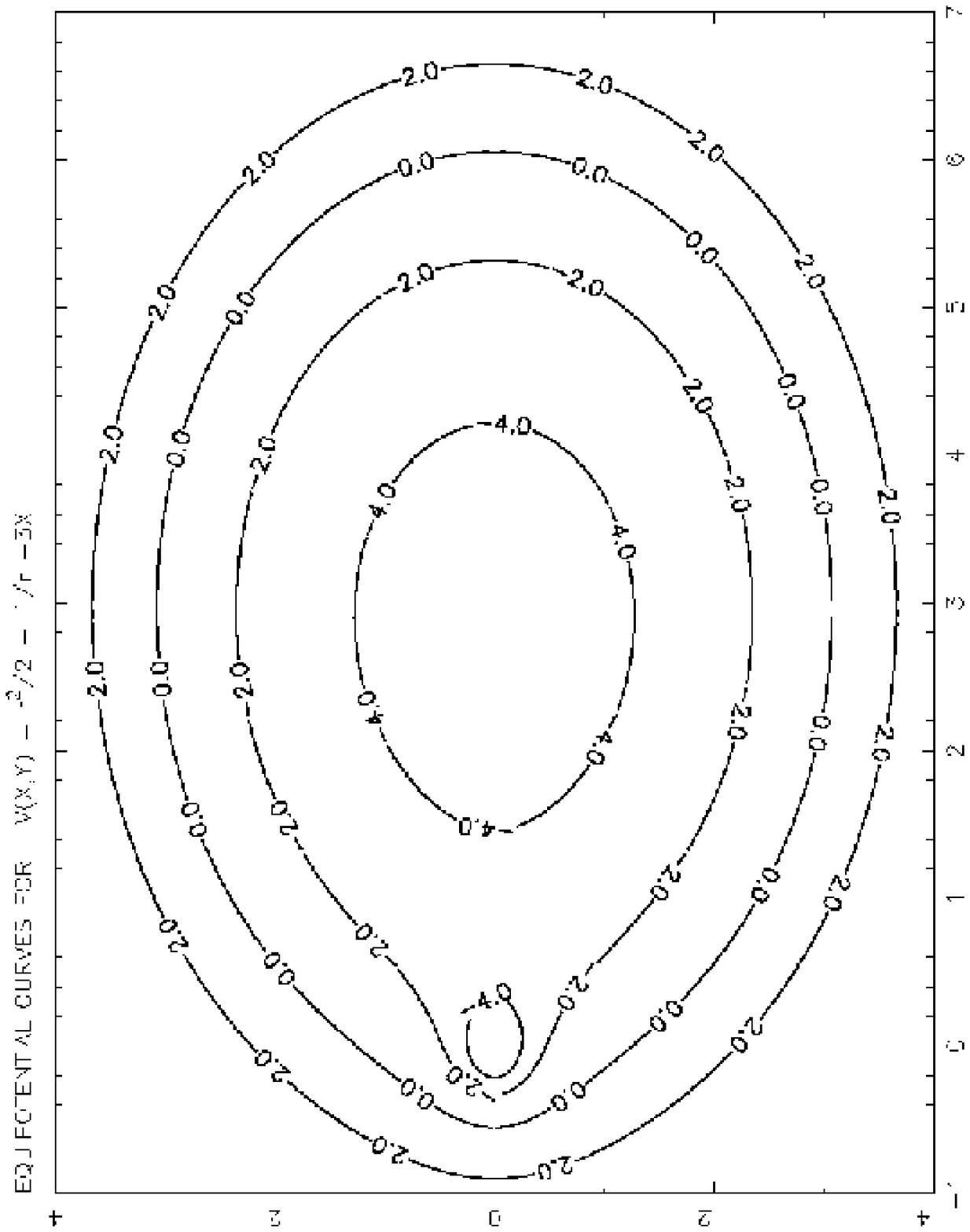}%
\caption{Equipotential curves for the two dimensional case with b=3  
\label{fig1}}
\end{figure}

Since energy is conserved phase space is three dimensional. The potential
energy \ $V=r^{2}/2-1/r-bx$ is singular at $\ r\rightarrow 0,$ and develops
a minimum when $b\geq 1.89$. \ref{fig1} shows equipotential lines for b=3.
Surface of section plots in the $x,p_{x}$ plane have been computed for
different values of the two parameters: b and the energy E.\ref{fig2a}shows
four plots, $E=0,b=1$ $and$ $b=3;E=-.5,b=1$ and $3.$ The plots appear
regular without any chaotic orbits. We have carried out many more
computations for different values of \ E and b with similar results. It
appears therefore that an additional constant of motion exists, but the
analytic expression has not been found. We have also carried out
computations to find the largest Lyapunov exponent which turned out to be
zero as expected for non-chaotic orbits.

Turning to the three dimensional case we study the $\boldsymbol{\alpha =}0$
limit. in this case it is convenient to introduce polar coordinates where
the system is described by the Hamiltonian

\begin{equation}\label{E:16}
H=P_{\rho }^{_{{}}^{2}}/2+P_{z}^{2}/2+P_{\varphi }^{2}/(2\rho ^{2})+\rho
^{2}/2-1/\sqrt{\rho ^{2}+z^{2}}
\end{equation}

where $\rho ^{2}=x^{2}+y^{2}$ , and \ $\varphi $ is an ignorable coordinate
so \ $P_{\varphi }=const.$ This gives the equations of motion

\begin{equation}\label{E:17}
  \begin{array}{l}
   \frac{\partial H}{\partial P_{\rho }}=\dot{\rho}\\[10pt]
   \frac{\partial H}{\partial P_{z}}=\dot{z}\\[10pt]
   \dot{P}_{\rho }=-\frac{\partial H}{\partial \rho }=-\rho +P_{\varphi
   }^{2}/\rho ^{3}-\frac{\rho }{(\rho ^{2}+z^{2})^{3/2}}\\[10pt]
   \dot{P}_{z}^{{}}=-\frac{\partial H}{\partial z}=-\frac{z}{(\rho
   ^{2}+z^{2})^{3/2}}
  \end{array}
\end{equation}

A surface of section plot ($\rho ,P_{\rho })$ in the z=0 plane is shown in
\ref{fig3} for E=-.5, $P_{\varphi }=.25$. The existence of chaotic orbits is
obvious, so the three dimensional equations of motion are not integrable.
To show that chaotic orbits exist \ for $\boldsymbol{\alpha \neq }0$, the
largest Lyapunov exponent has been computed for the three dimensional case
for b=3, as shown in \ref{fig4}, using the algorithm as described in Ref.\cite{eO93}
It converges to a value larger then zero as expected.\\

In conclusion the motion of positronium immersed in a magnetic field is
chaotic in the classical limit, therefore the long lifetime predicted in the
quantum limit is unlikely.

\newcounter{mycount}
\renewcommand{\thefigure}{\arabic{mycount}.\alph{figure}}
\setcounter{mycount}{2}
\setcounter{figure}{0} 

\begin{figure}
\vspace{1.in}
\includegraphics{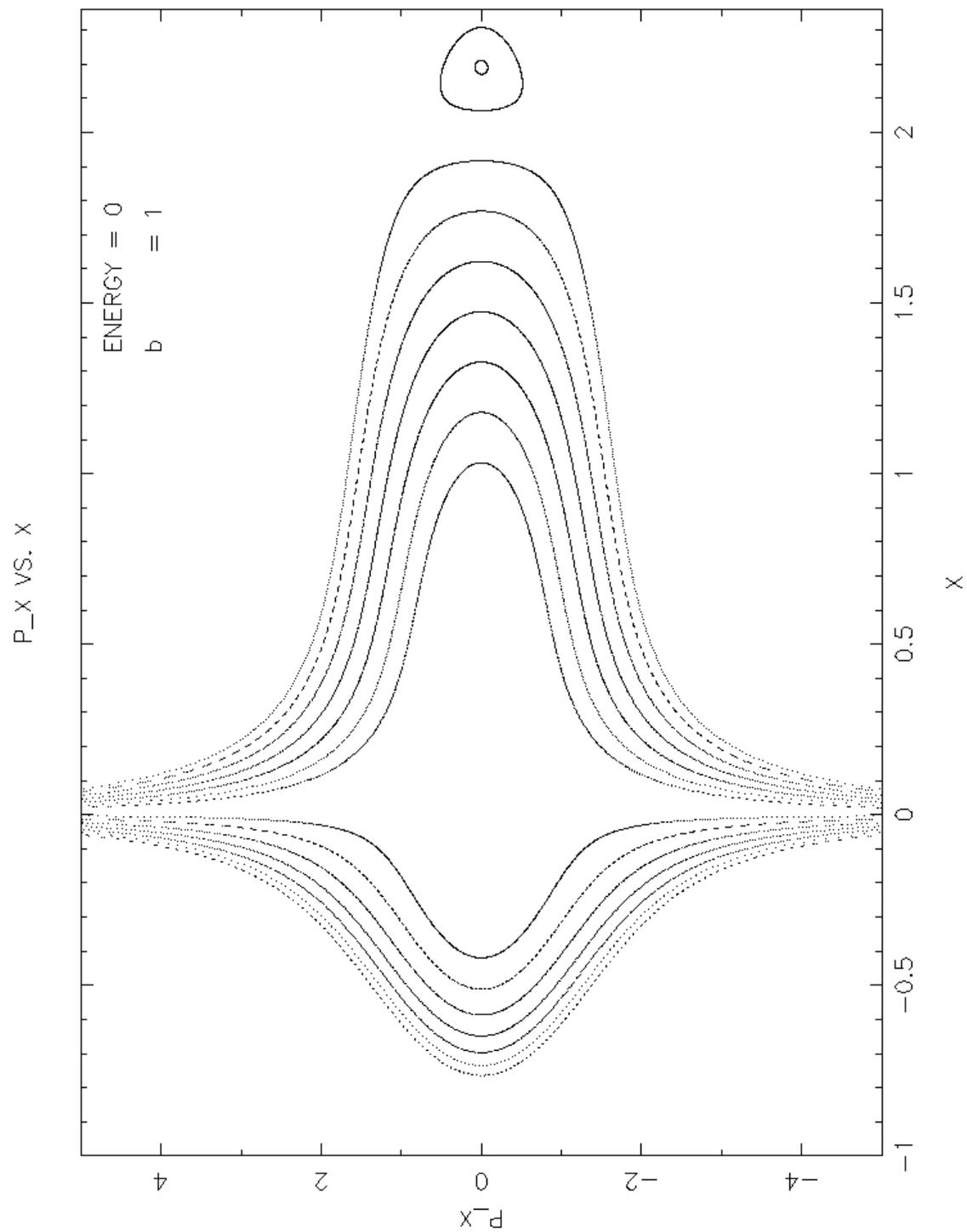}%
\caption{Surface of section plots in the $x-p_{x}$ plane,E=0,b=1
\label{fig2a}}
\end{figure}

\begin{figure}[tbh]
\vspace{1.in}
\includegraphics{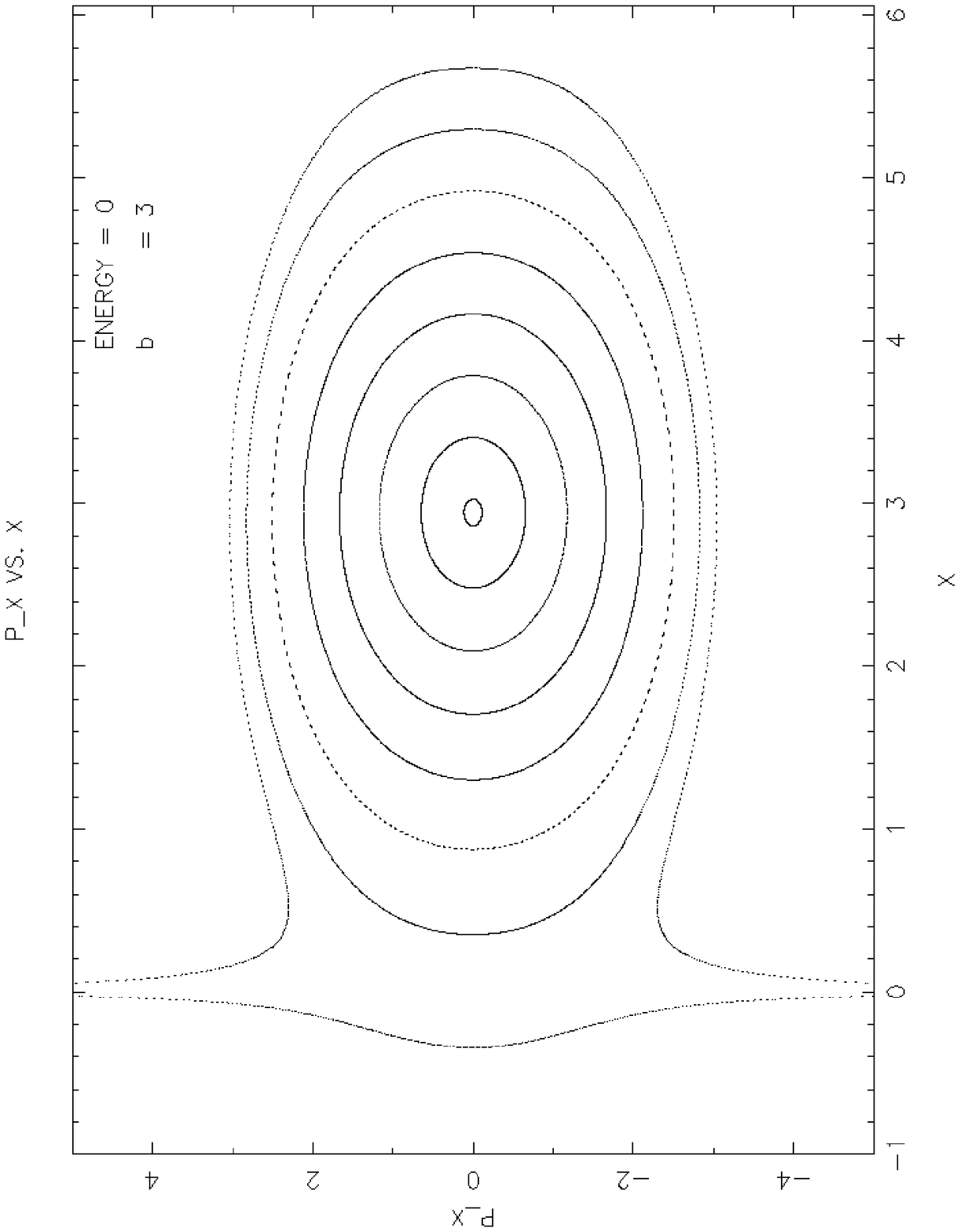}%
\caption{E =0,b = 3\label{fig2b}}
\end{figure}

\begin{figure}[tbh]
\vspace{1.in}
\includegraphics{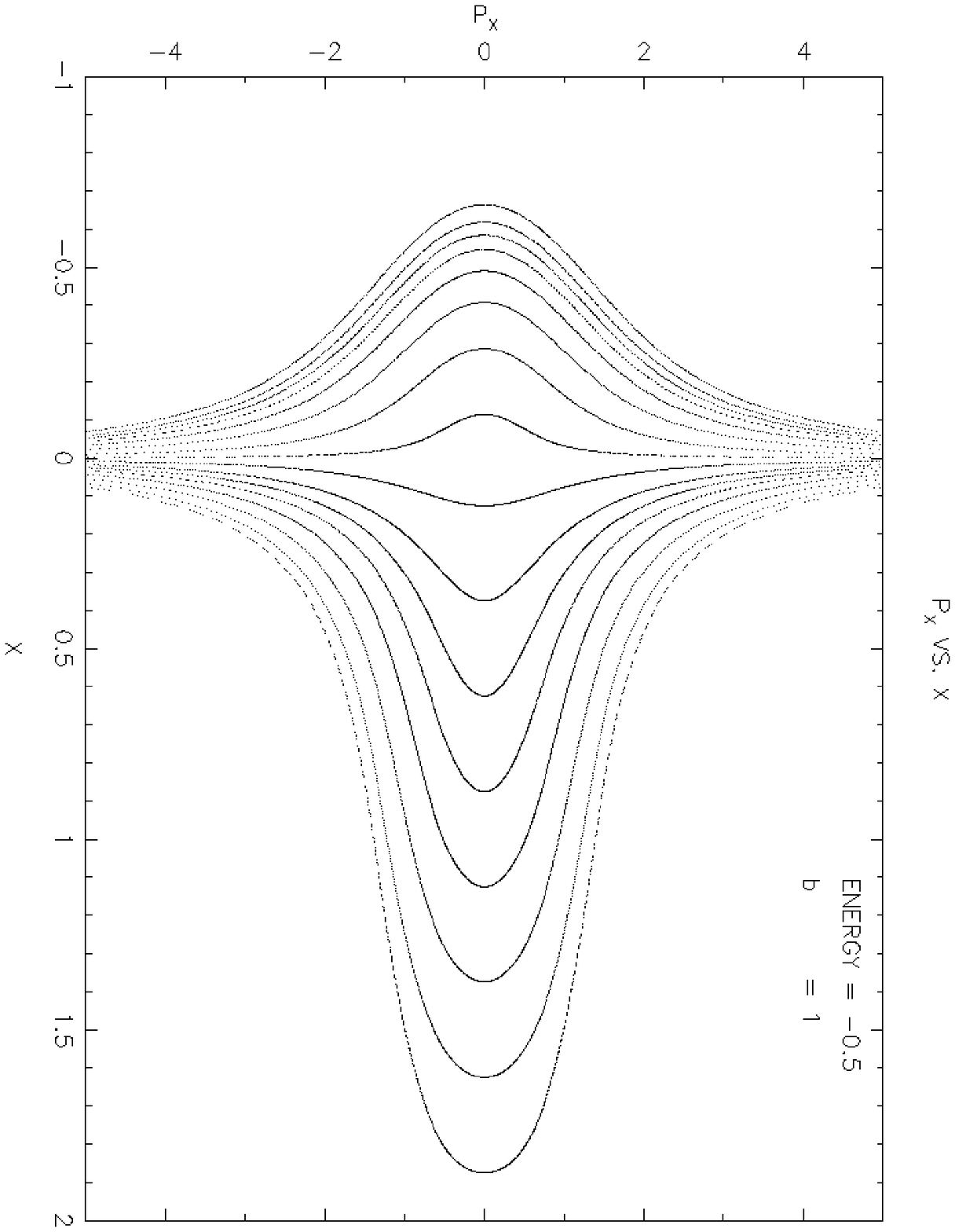}%
\caption{E = -0.5, b = 1 \label{fig2c}}
\end{figure}

\begin{figure}[tbh]
\vspace{1.in}
\includegraphics{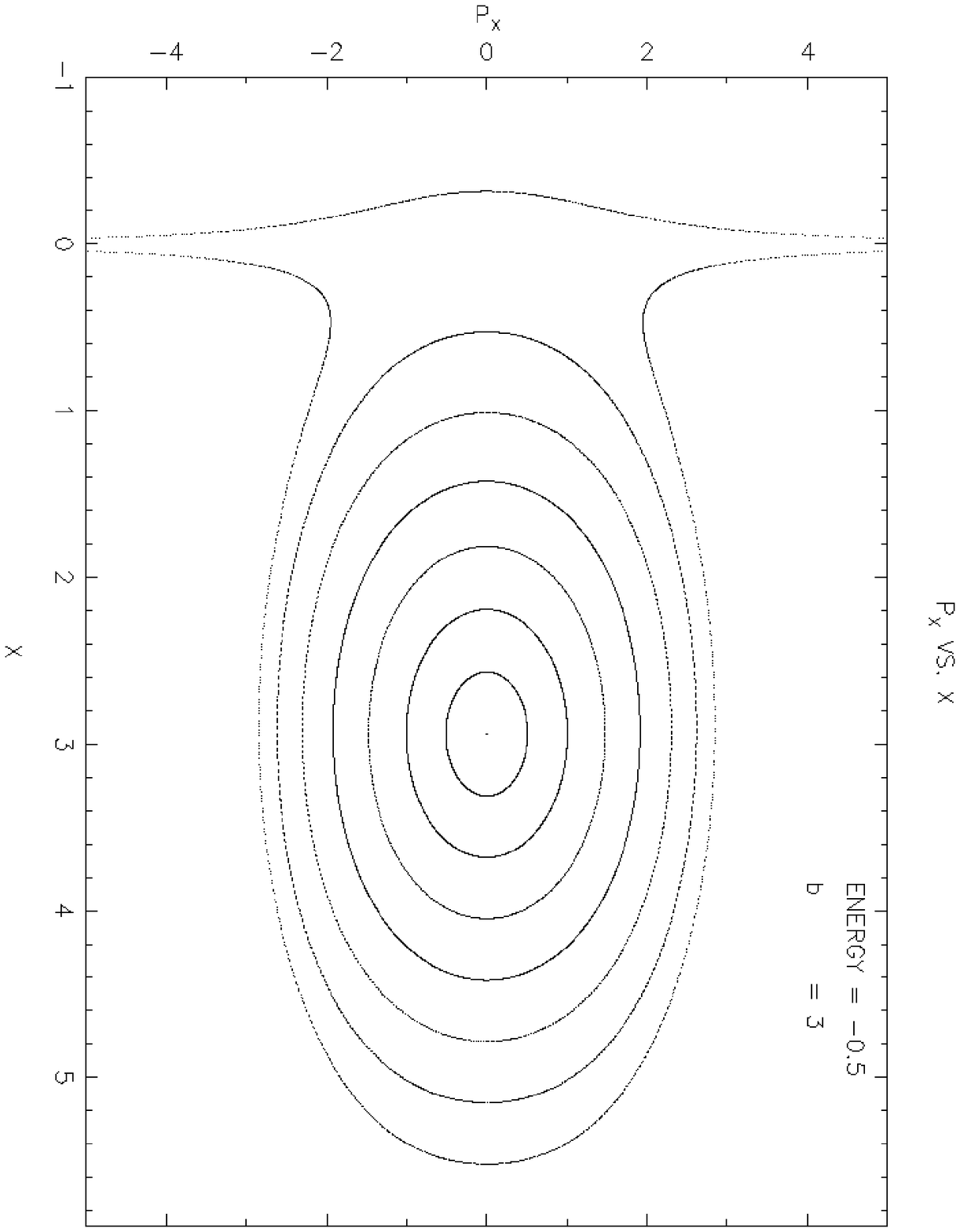}%
\caption{E = -0.5, b = 3\label{fig2d}}
\end{figure}

\renewcommand{\thefigure}{\arabic{figure}}
\setcounter{figure}{2}

\begin{figure}[tbh]
\vspace{1.in}
\includegraphics{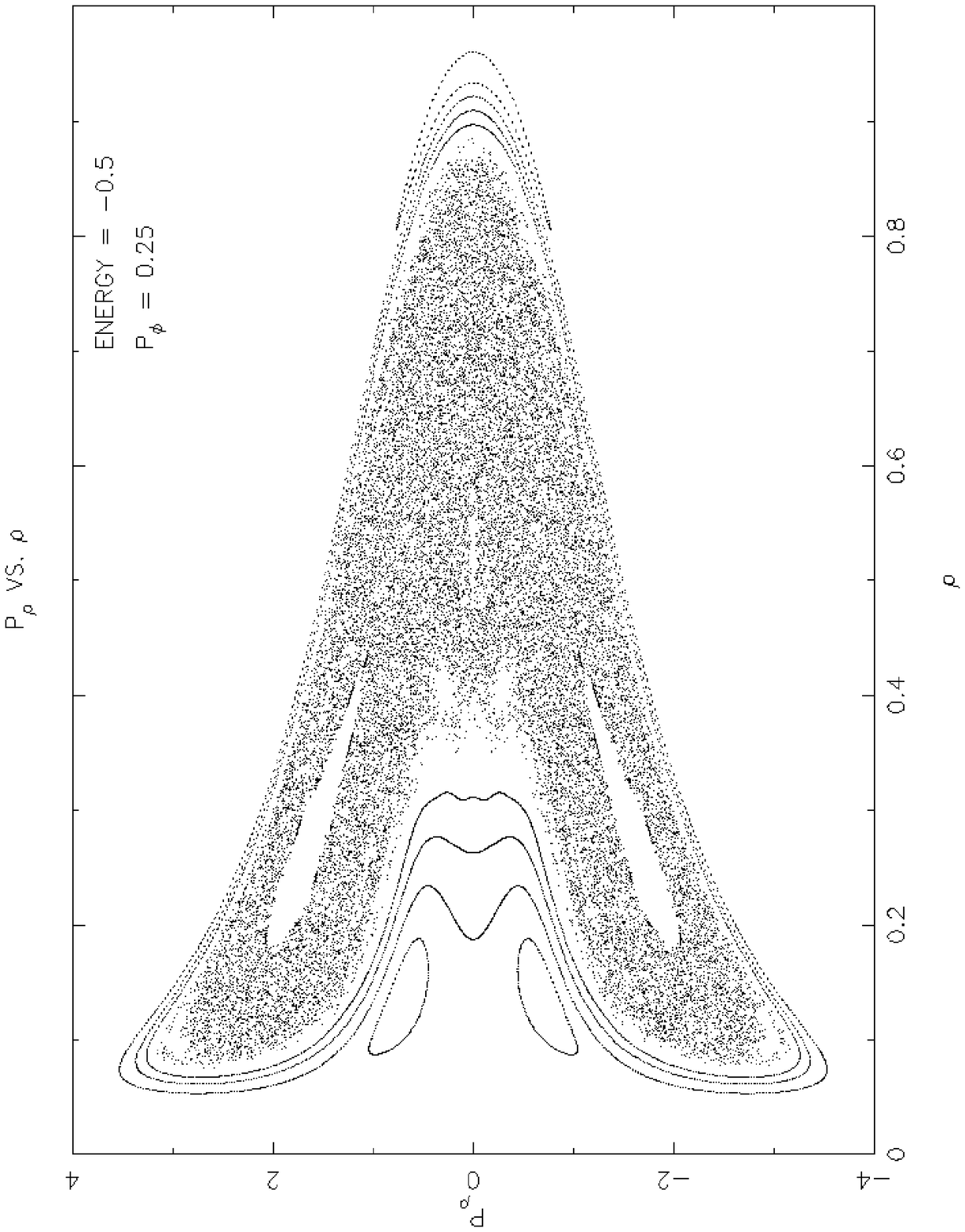}%
\caption{ Surface of sections plot for the three dimensional case, where b= 0
in the $\rho -P_{\rho }$ plane\label{fig3}}
\end{figure}

\begin{figure}[tbh]
\vspace{0.25 in}
\includegraphics{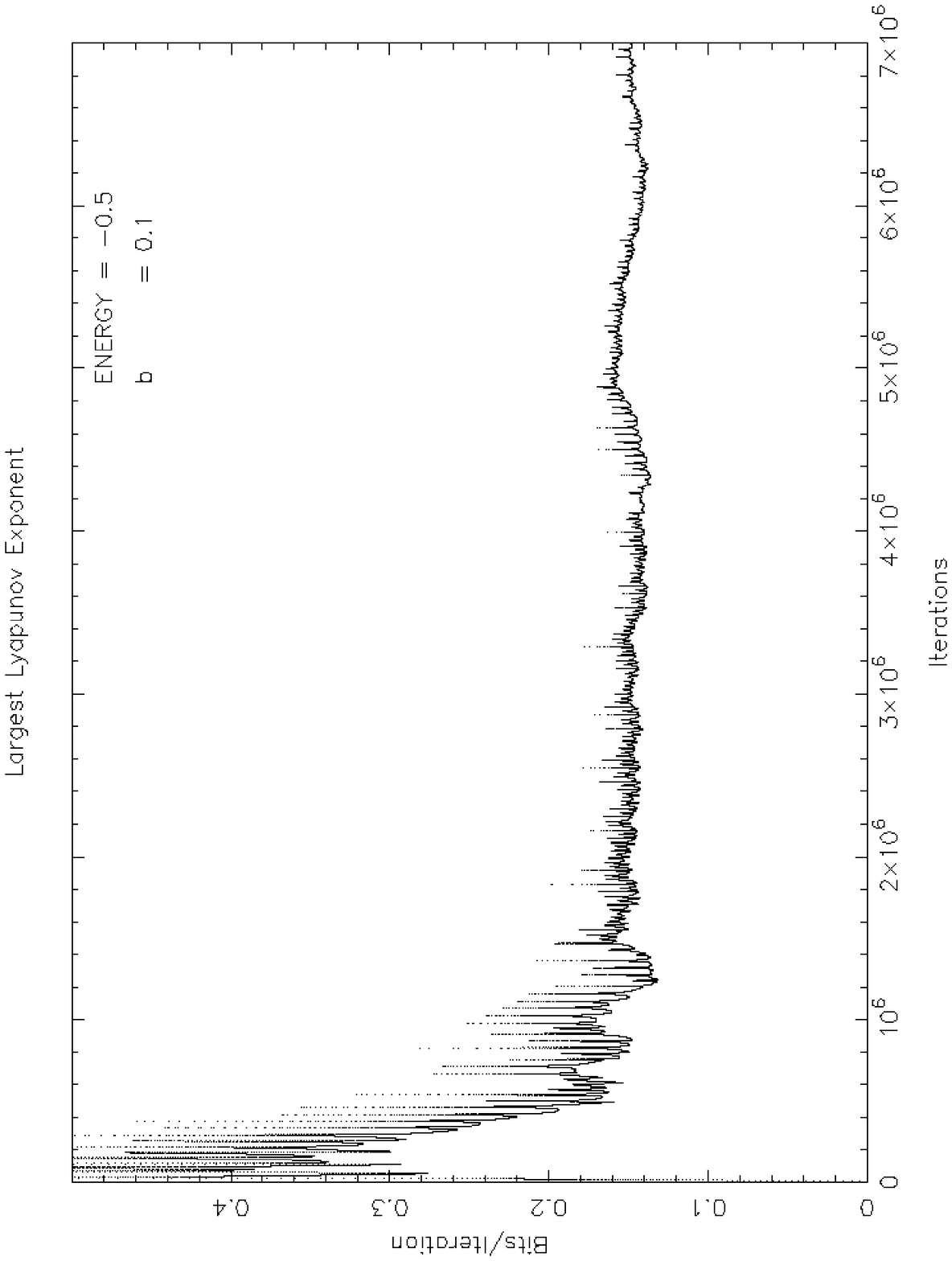}%
\caption{Computation of the largest Lyapunov exponent, E=-0.5, b=0.1
with initial conditions\\ 
$x_1$ = 0.65,$ y_1$ = 0.0,$ z_1$ = 0.0\\ 
$P_{x1}$ = 0.0,$ P_{y1}$ = 0.25,$ P_{z1}$ = 1.31222067 \\
$ x_2$ = 0.65000001,$ y_2$ = 0.0,$ z_2$ = 0.0\\ 
$P_{x2}$ = 0.0,$ P_{y2}$ = 0.25,$ P_{z2}$ = 1.31222064
\label{fig4}}
\end{figure}

\end{document}